# The MoE-Empowered Edge LLMs Deployment: Architecture, Challenges, and Opportunities

Ning Li, Song Guo, *Fellow, IEEE*, Tuo Zhang, Muqing Li, Zicong Hong, Qihua Zhou, Xin Yuan, Haijun Zhang, *Fellow, IEEE*

*Abstract*—The powerfulness of LLMs indicates that deploying various LLMs with different scales and architectures on end, edge, and cloud to satisfy different requirements and adaptive heterogeneous hardware is the critical way to achieve ubiquitous intelligence for 6G. However, the massive parameter scale of LLMs poses significant challenges in deploying them on edge devices due to high computational and storage demands. Considering that the sparse activation in Mixture of Experts (MoE) is effective on scalable and dynamic allocation of computational and communications resources at the edge, this paper proposes a novel MoE-empowered collaborative deployment framework for edge LLMs, denoted as CoEL. This framework fully leverages the properties of MoE architecture and encompasses four key aspects: Perception, Deployment, Compression, and Updating. Edge servers broadcast their resource status and the specific resource requirements of LLMs to their neighbors. Then, utilizing this data, two sophisticated deployment strategies are proposed for satisfying varying model scales, ensuring that each model is deployed effectively. One for deploying LLMs on a single edge device through intra-device resource collaboration, and another for a distributed deployment across multiple edge devices via inter-device resource collaboration. Furthermore, both the models and the intermediate data are compressed for reducing memory footprint by quantization and reducing the volume of intermediate data by token fusion and pruning. Finally, given the dynamic of network topology, resource status, and user requirements, the deployment strategies are regularly updated to maintain its relevance and effectiveness. This paper also delineates the challenges and potential research directions for the deployment of edge LLMs.

## I. INTRODUCTION

The proliferation of large language models (LLMs), driven by the transformative success of transformers, has captured the attention of the AI community and beyond. Nowadays, leading players in the AI industry are competing to develop their own LLMs, such as GPT-4 [1], LLaMA [2], Mixtral 8 × 7B [3], etc. These models demonstrate a remarkable phenomenon known as "emergence", where their generalization capabilities are significantly enhanced as their model-size grows. These exceptional capabilities allow them to be directly applied or easily adapted to a multitude of downstream and unseen tasks, unlocking vast potential across various applications, including Chatbot, content generation, and healthcare [4]. The powerfulness of LLMs indicates that deploying various LLMs with different scales and architectures on end, edge, and cloud to satisfy different requirements and adaptive heterogeneous hardware is the critical way to achieve ubiquitous intelligence for 6G.

However, the enormous parameter scale of current LLMs necessitates vast computational and storage resources for both training and inference, making them heavily reliant on cloud data center. This reliance poses significant challenges for deploying LLMs on end and edge devices, which typically have limited resources. Solely depending on cloud-based LLMs is insufficient for providing efficient and convenient ubiquitous intelligence for 6G, as they encounter several unexpected issues, including latency in real-time applications, substantial bandwidth consumption due to large data transmissions, and significant privacy concerns when handling sensitive data in cloud environments [5].

In response to these challenges, several studies have proposed end LLMs [6]. These approaches utilize model and knowledge compression techniques to reduce the model size, making them suitable for resource-constrained edge devices while striving to preserve the performance of LLMs. However, since certain empirical scaling laws reveal a power-law relationship between the final model capability and model scale [9], the accuracy loss and capability loss are inevitable compared to the larger-scale cloud-based LLMs. Thus, it is increasingly clear that deploying larger models than end LLMs at the edge and cooperating with end LLMs will be the promising solution to achieve high-effective ubiquitous intelligence for 6G. This approach is emerging as a key focus for future innovation of LLMs, aiming to balance the need for scalability with the constraints of edge environments.

However, given the constraints of limited, heterogeneous, and dynamic resources at the edge, research into effectively deploying LLMs on edge devices is still in its infancy, and high-efficiency edge LLMs continues to confront significant obstacles and challenges [6][13]. Recently, according to the different architecture properties of LLMs, i.e., dense model [4][5] or MoE [7][8], several studies have suggested leveraging edge collaboration for deploying LLMs at the edge. However, the dense models feed all parameters to each input token, which is not effective considers the constraints of edge environment. Different with the dense model, a typical MoE architecture is sparse. It consists of a gated network and several expert networks, and selectively activates a portion of parameters for various inputs to participate in computation. Through its sparsity design, the MoE is able to scale its parameter size and capability with almost constant computing complexity, making them good fit to edge devices. The sparse activation in MoE architecture is much more effective on scalable and dynamic allocation of computational and communications resources than the dense model [7][8]. Thus, fully leveraging the architecture properties of MoE is an invaluable solution for edge LLMs.

Nevertheless, setting aside the fact that edge LLMs i are still in their nascent stages, the deployment of LLMs on the edge using the MoE architecture also confronts numerous challenges. *Firstly*, the MoE architecture tends to be more expansive than its dense counterparts with equivalent capabilities, making it challenging to fit within the memory constraints of edge devices. Simply scaling down the expert number or compressing the model size could significantly degrade the performance capacity [6]. What's worse, as each edge server may need to support multiple LLMs for diverse tasks and applications, the scarcity of edge resources is likely to become even more critical. *Secondly*, the frequent All-to-All communication delays inherent in MoE models are significantly more pronounced compared to those in their dense counterparts. Additionally, the collaborative edge server operations, which necessitate substantial data transmission across different edge servers, make this situation even worse. Because the data transmission across edge servers is via internet connections, despite the high bandwidth of optical fiber, its transmission capability remains inferior to internal I/O speeds within a device, leading to significant delays in Edge LLM inference. *Finally*, on the one hand, the capabilities and resources of edge servers, along with the varying demands of users and applications, are both heterogeneous and dynamic; on the other hand, the resource collaborative in edge network is multidimensional, i.e., the inter-server resource collaborative between different servers and the intra-server resource collaborative between GPU memory and SSD. These diversities substantially reduce the efficiency of aligning the MoE expert allocation strategy with the heterogeneous and dynamic resource and user demands, all while striving to maintain high inference performance. Therefore, a more sophisticated experts allocation strategies is required to optimize the performance of edge LLMs.

Considering the aforementioned challenges, this paper introduces a novel collaborative deployment framework for edge LLMs based on the architecture properties of MoE, denoted as CoEL. This framework is flexible, multi-dimensions resource collaborative, dynamic self-adaptive, and low communication cost. The CoEL encompasses four key components: Perception, Deployment, Compression, and Updating. In the perception phase, edge servers broadcast their resource status and the specific resource requirements of LLMs to their neighbors. Then, utilizing this data, an optimal deployment strategy is formulated for satisfying varying model scales, ensuring that each model is deployed effectively. In the CoEL framework, two sophisticated deployment strategies are proposed: one for deploying LLMs on a single edge device through intra-device resource collaboration, and another for a distributed deployment across multiple edge devices via inter-device resource collaboration. Additionally, the distributed deployment strategy considers the frequency of intermediate data transmissions between devices. During the compression phase, both the models and the intermediate data are compressed for: 1) reducing memory footprint by quantization and 2) reducing the volume of intermediate data by token fusion and pruning. Finally, given the dynamic nature of network topology, resource status, and user requirements, the deployment strategy must be regularly updated and adjusted to maintain its relevance and effectiveness. Furthermore, this paper delineates the challenges and potential research directions for the deployment of edge LLMs.

## II. CONVENTIONAL LLM DEPLOYMENT STRATEGY AND ITS LIMITATIONS FOR EDGE LLMS

### A. Overview of Conventional LLM Deployment Paradigms

**Deployment on cloud**. This is the primary deployment solution for large-scale LLMs [9], such as Mixtral 8×7B, ChatGPT, and LLaMA. Despite the considerable capabilities of cloud infrastructure, it is still impractical to deploy LLM on one single GPU. Large cloud data centers, such as those operated by AWS and Azure, typically contain thousands of servers, with each server equipped with 4 or 8 GPUs to meet the high demands of large-scale LLMs. Consequently, these LLMs are segmented into smaller sub-models and distributed deployed across various GPUs, either within the same server or across different servers within the same data center, employing techniques like expert parallelism, tensor parallelism, pipeline parallelism, model parallelism, etc. [9]. The inference tasks are processed on each device and then synchronized through high-speed communication technologies, including PCIe, NVlink, InfiniBand [10], and RDMA [11], ensuring high-performance distributed and collaborative inference within the cloud data center. The details of the cloud-based deployment are presented in Table I.

**Deployment on end devices**. On the one hand, due to the rapid advancements in hardware, the computing and storage capability of end devices becomes powerful, making it feasible to deploy appropriate scale of LLMs directly on end devices. On the other hand, by leveraging high-efficiency model architectures such as MoE and parameter sharing, along with compression techniques including quantization, pruning, knowledge distillation, low-rank decomposition [6], etc., LLMs can be effectively deployed on end devices. For example, models like Llama-7B can be deployed on devices like the Xiaomi 12 using 8-bit quantification [12]. These deployment strategies demonstrate the potential for on-device LLM deployment. However, given the still limited capacity of end device, the scale of LLMs that can be effectively deployed on it remains constrained, such as smaller than 10B [6]. The details of the end-based deployment are presented in Table I.

**Deployment on cooperated edge**. The edge collaborative deployment for LLMs bridges the performance gap between cloud LLMs and end LLMs. The collaborative and distributed deployment can divide the computational workload of LLM inference (e.g., tensor parallelism, pipeline parallelism, etc.) across various edge servers, enabling real-time processing of LLMs even with limited capabilities [13][14]. Researchers have developed several strategies to optimize LLM throughput, latency, and resource utilization across edge servers. For example, EdgeShard [4], Galaxy [5], Edge-MoE [7], WDMoE [8], etc. These innovative approaches enable a flexible trade-off between response quality and resource consumption, making them suitable for edge scenarios. Furthermore, edge LLMs face more pronounced heterogeneity and dynamism in resource availability and user demands compared to cloud LLMs and end LLMs, leading to increased deployment complexity [6][13]. These factors distinguish deploying LLMs at the edge from the practices of partitioning LLMs and distributing them across multiple GPUs in cloud data centers. The details of the edge-based deployment are presented in Table I.

Table I. The compression of different deployment strategies

| | Deployment | Parallelism | Heterogeneous | Dynamic | Deployment Complex | Data Transmission Method | | Technical Maturity |
|---|---|---|---|---|---|---|---|---|
| Cloud LLMs | Multi-Devices & Multi-GPUs | ➤ Pipeline Parallelism<br>➤ Tensor Parallelism<br>➤ Expert Parallelism<br>➤ Model Parallelism | Weak | Weak | ★☆☆ | In-Device | PCIe NVLink | ⏱ |
| | | | | | | Between-Devices | InfiniBand RDMA | |
| End LLMs | Single-Device & Single-GPU | NoN | Weak | Weak | ★★☆ | PCIe | | ◐ |
| | Single-Device & Single-NPU | | | | | | | |
| Edge LLMs | Single-Device & Single-GPU | NoN | Strong | Strong | ★★★ | In-Device | PCIe NVLink | ⏱ |
| | Single-Device & Multi-GPUs | ➤ Pipeline Parallelism<br>➤ Tensor Parallelism<br>➤ Expert Parallelism<br>➤ Model Parallelism | | | | Between-Devices | LAN | |
| | Multi-Devices & Multi-GPUs | | | | | | | |

*B. Disadvantages of Current Paradigms for Edge LLM*

The various deployment strategies mentioned have indeed enhanced the performance of LLMs across diverse application scenarios, particularly for end deployment and edge collaborative deployment, making it feasible to deploy LLMs at end and edge. Nonetheless, applying these strategies on edge LLMs still encounters certain disadvantages considering the limited, dynamic, and heterogeneous edge environment.

**Cloud LLMs**. For the cloud LLMs, the models are also deployed in a distribution and collaborative manner across various GPUs in the same server or across different servers within the same data center, as referenced in [4][9]. However, this strategy is not effective in edge LLMs, as it neglects the heterogeneous and resource constrained edge computing scenario. First, cloud servers are typically equipped with homogeneous GPUs, whereas edge devices inherently possess heterogeneous computational capabilities. Second, the cloud GPUs used for LLMs are connected by high-bandwidth links such as InfiniBand and NVLinks, which can achieve speeds up to 600 GB/s; whereas edge servers are often connected by the links with bandwidth ranging from tens of Kbps to 1000Mbps. This disparity in connectivity significantly impacts the effectiveness of deploying collaborative and distributed strategies from cloud environments to edge settings.

**End LLMs**. The end LLMs can adaptive the limited and heterogeneous resource in end devices effectively. Nonetheless, as dictated by the scaling law, the intelligence performance of large-scale LLMs consistently outperforms that of small-scale LLMs [9][12]. Unfortunately, to accommodate the limited resource in end devices, the models in end LLMs are compressed based on pruning, quantization, knowledge distillation, low-rank compression, etc. Therefore, the accuracy and capability loss are inevitable compared to the larger-scale cloud LLMs.

**Edge LLMs**. A primary limitation of current edge collaborative solutions is that they fail to fully leverage the properties of various model architectures, such as the sparse activation inherent in MoE, to ensure high-performance edge LLMs. Moreover, to mitigate the delays caused by low-bandwidth communication link, it is crucial to reduce not only the volume of transmitted data but also the frequency of transmissions. For example, by strategically deploying different MoE experts on suitable edge servers to ensure that the majority of inference tasks are processed locally, the delay caused by intermediate data transmission can be significantly minimized. Unfortunately, these critical aspects are overlooked in both cloud-based and edge-based deployment strategies.

## IV. PROPOSED APPROACH

According to the aforementioned challenges, this paper introduces CoEL framework for the deployment of LLMs at the edge. The CoEL framework is flexible, multi-dimensions resource collaborative, dynamic self-adaptive, and low communication cost. The flexibility of the CoEL framework allows it to effectively accommodate various model scales. The multi-dimensions resource collaboration within CoEL encompasses both inter-server and intra-server resource coordination. The dynamic self-adaptive enables the framework to dynamically adjust and refine deployment and compression strategies in accordance with resource availability and user demands. The low communication cost in CoEL is achieved by reducing both the volume and frequence of intermediate data transmission in collaborative model deployment. The CoEL framework is composed of four essential components: Perception, Deployment, Compression, and Updating. Further details of the CoEL are elaborated upon in the following sections.

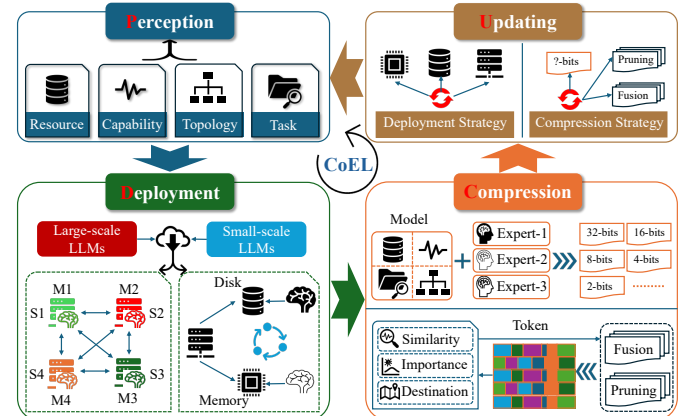

Fig.1. The main procedures of the proposed CoEL framework

*A. Approach Overview*

**Information Perception.** Prior to deploying LLMs on edge servers, it is crucial to determine the cooperative strategy, i.e., determining the number of edge servers to be involved and which specific servers to select, based on the resource status of edge servers and the resource demands of tasks. Consequently, the information regarding edge servers and the models to be deployed must be regularly perceived and collected. This is the fundamental for deployment. Specifically, as shown in Fig.1, the edge servers exchange information with its neighbors periodically. The information includes the status of available resources (the GPU and CPU memory, storage of SSD), the computing capability, the network topology, the scale of models, etc. The information is exchanged by the Hello message, which is periodically broadcast to all surrounding servers during runtime. The information of server's available resources, computing capability, and resource requirements of model is embedded into the Hello message without adding extra bits. For instance, within the 96-bits OLSR hello message, there is a reserved space of 16 bits, and 8 bits are sufficient to express the percentage of available computational or memory resources, thus utilizing 16 bits in total. This information is transmitted only when changes in server resources exceed a predefined threshold, ensuring efficient communication without unnecessary updates.

**Flexible Model Deployment.** Deploying various LLMs with different scales and architectures on end, edge, and cloud to satisfy different requirements and adaptive heterogeneous hardware is essential for realizing the vision of ubiquitous intelligence in 6G. Consequently, considering the diverse scales of models and the capabilities of devices, deployment strategies must be customized for different models to optimize performance and resource utilization. For instance, if the model is small, e.g., smaller than 10B, it is possible to deploy it on single edge server [12]; otherwise, a collaborative deployment across multiple edge servers becomes necessary [9]. Furthermore, even with the same model, the deployment strategies vary depending on the capabilities of edge servers. Consequently, leveraging the perception data, we propose two collaborative expert deployment strategies: one for inter-device resource cooperation and another for intra-device resource cooperation.

As illustrated in Fig.1, for the inter-devices resource cooperation, based on the resource status of edge servers and the resource requirements of LLM, the optimal edge server collaboration region is calculated. Then, according to the connection probabilities between different experts in different expert layers, the expert layers in LLM are horizontally segmented into a series of sub-models. Each sub-model has the same number of layers as the original model but varies in the number of experts per layer. For the experts in the adjacent expert-layers, the higher the connection probability between them, the more likely they are to be included in the same sub-model. Additionally, experts in the same layer can be repeated across different sub-models. Then, a resource-aware gate network is trained to route the tokens to the optimized expert when this expert is repeatedly deployed on different edge servers. Subsequently, according to the resource status of edge server and the scale of model, each server then decides whether to deploy a subset of the sub-models or the entire set, aiming to minimize the inference delay and the frequency of intermediate transmission across servers. Certainly, the sub-models that deployed between the collaborative edge servers should encompassed all the experts from the original model. According to Fig.1, Fig.2 illustrate how this strategy works by a toy example.

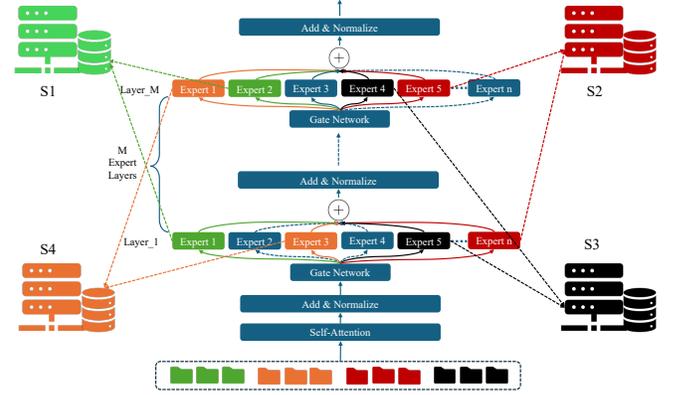

Fig.2. A toy example of how the inter-devices resource cooperation based deployment strategy works. (In this example, different color represents different expert popularity under different tokens from different servers)

For the intra-device resource cooperation, as shown in Fig.1, according to the sparse activation characteristic of MoE, the experts are dynamically scheduled between memory (GPU, CPU) and SSD to reduce the memory footprint. Specifically, the weights of non-expert layer are entirely loaded into the GPU memory. Then, based on the distribution of Token requests, a small model is trained to predict the expert popularity in different expert layers. When the Tokens are processed in the shallow layers, the small model predicts the experts to be activated in deep layers in advance and preload them into the I/O pipeline. Meanwhile, synchronously updating the expert scheduling strategy in deep expert layers between GPU-memory and SSD, ensuring efficient resource utilization. The experts scheduling strategy can be utilized in deploying the LLMs on one single edge server or in conjunction with inter-devices resource cooperation.

**Model and Token Compression.** Even the inter-device and intra-device resource collaborative improve the deployment efficiency of edge LLMs, it still faces the following disadvantages: 1) since various LLMs (with different scales and for different applications) need to be deployed on edge to achieve ubiquitous intelligence, the limited, heterogeneous, and dynamic of resource in edge servers becomes even more serious; 2) for the distributed deployment strategy in CoEL, as the bandwidth is limited, the intermediate data transmission should be reduced further. The model compression and token compression are proposed in CoEL to address these issues.

Regarding to model compression, as illustrated in Fig.1, a dynamic and self-adaptive expert quantification strategy for edge LLMs is proposed to reduce the memory footprint. This strategy employs bit-width adaptation and mixed-precision quantization, allowing the weights and activation values of experts to be quantized individually based on the available resources of servers and the requirements of models. The goal is to optimize inference accuracy while minimizing inference latency. Specifically, a small model analyzes the distribution of Tokens and predicts the popularity of experts under varying resource conditions and application demands. For the less popular experts, a lower bit quantization is utilized, such as 4-bits or 8-bits precision, whereas more pupular experts are assigned either full precision or higher bit quantization, like 16-bits. Meanwhile, the weights of non-expert layers are quantized by the same self-adaptive and personalized quantization technique, ensuring an optimized balance between inference accuracy and computational efficiency.

The distributed deployment of LLMs across edge devices requires substantial inter-server communication for the intermediate data transmission. Although the frequency of inter-device transmission has been reduced adequately during deployment, given the limited, unstable, and dynamic of bandwidth between edge servers, we enhance the transmission efficiency further by reducing the volume of intermediate data. Specifically, as shown in Fig.1, based on the distribution of Token requests, a small model is utilized to predict the optimal experts for each Token. Once the optimal experts are not deployed on the current edge server, the following measures are taken to reduce the volume of data transmitted between edge servers. On the one hand, the tokens which utilize the same optimal expert in the same edge server are fused based on the similarity before inputting into the expert. On the other hand, the tokens which have little effect on the final inference result will be pruned.

**Dynamic Deployment Updating.** On the one hand, since the expert popularity varies with the server's resource status and user's token requests, the model deployment and compression strategies need to be updated dynamically. On the other hand, the parameters of LLMs need to be updated with the system operating. Specifically, as shown in Fig.1, the expert deployment strategies, i.e., on which edge servers, in memory or SSD, etc., adjust according to the variation of the servers' resource status and the users' requirements. Then, since the low precision (e.g., 4bits) experts cannot retrieve to high precision, when the experts' parameters need to be retrieved to higher precision (e.g., 16bits), the edge server searches the high precision parameters of these experts from the collaborative servers and quantizes them to the required bit-width. Otherwise, the edge server downloads the full precision parameters of these experts from cloud. The purpose is to minimize the latency for parameter transmission. Additionally, as the system operating, model parameters need to be continuously updated to adapt to new data and environment, ensuring the accuracy and effectiveness of model. Consequently, in CoEL framework, the entire model is fine-tuned at edge servers based on local data, following the principles of incremental learning technology.

### B. Advantages and Novelty

**Flexible model deployment strategy based on multi-dimensions resource cooperation.** To realize ubiquitous intelligence for 6G, various LLMs (with different scales and for different applications) need to be deployed at the edge. Considering the heterogeneity in model sizes, resource capabilities, and user requirements, this paper proposes two deployment strategies based on inter-device resource cooperation and intra-device resource cooperation, to offer flexible deployment options for edge LLMs.

**High-effective and dynamic deployment strategy by fully utilizing the model architecture.** Distinguished from previous approaches, the CoEL framework leverages the architecture properties of MoE to address the constraints of limited, dynamic, and heterogeneous edge resources. Within CoEL, by meticulously and adaptively distributing experts across different edge servers, or scheduling their allocation between GPU memory and SSD, both the memory footprint and the intermediate data transmission (both volume and frequency) are reduced effectively, all while sustaining high inference performance of edge LLMs.

**Dynamic self-adaptive model and transmission compression strategy.** To further accommodate the limited, dynamic, and heterogeneous of edge resources, this paper integrates mixed-precision quantization technology into the MoE architecture, enabling dynamic bit-width adaptation for different experts. Furthermore, the CoEL framework proposes the dynamic token fusion and pruning mechanism which aimed at reducing the volume of intermediate data transmission. This mechanism, in conjunction with flexible and adaptive deployment strategies, effectively minimizes both the data transmission volume and frequency.

## IV. PERFORMANCE EVALUATION

In this section, we conduct a proof-of-concept evaluation to substantiate the feasibility and effectiveness of the proposed CoEL framework. A prototype system has been designed based on the proposed CoEL framework and the LLM is deployed on a set of high capability edge servers under various distributed deployment approaches.

### A. Experiment Setup

Hardware and Network. As depicted in Fig.3, the network comprises three edge servers, configured in a star topology for interconnection. The server 1 equips with: GPU: 2*RTX 4090D, CPU: Intel(R) Core(TM) i9-14900KF, Host Memory: 64G; the hardware in server 2 and server 3 are the same: GPU: 1*RTX 4090D, CPU: Intel(R) Core(TM) i9-14900KF, Host Memory: 32G. The communication across different servers is LAN with the bandwidth is 1GB; the communication within the server is PCIe with the bandwidth is 64GB.

Model. The LLM model used in this experiment is Qwen1.5-MoE-A2.7B-Chat, which is the decode-only MoE model based on Transformer framework. The parameter scale is 14.3B. Additionally, the inference framework used in this experiment is vLLM.

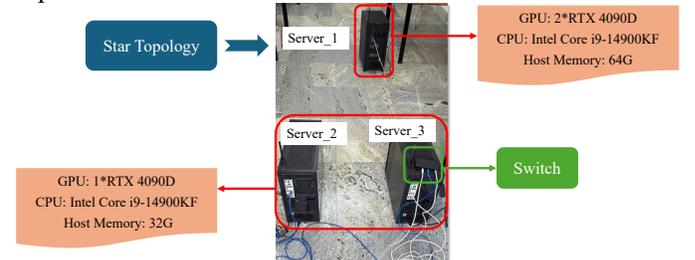

Fig.3. Hardware and Network

### B. Effectiveness of the proposed CoEL framework

The experimental results are displayed in Fig.4 to Fig.6. Fig.4 and Fig.5 illustrate the average generation throughput and average latency across varying output token lengths, ranging from 64 to 2048, with a fixed input token length of 128. Fig.6 shows the average latency under different input token lengths, with a constant output token length set at 128.

Fig.4 compares the average generation throughput under two distinct scenarios: 1) a single server equipped with two GPUs, and 2) two servers, each with one GPU. As demonstrated in Fig.4, the average generation throughput increases with the output token length, reaching a peak when the output token length exceeds 1024. Furthermore, the average generation throughput for the single server configuration is at most 1.7 times higher than that of the two distributed servers. In Fig.5, we compare the average latency under the above two scenarios. As shown in Fig.5, on the one hand, with the increasing of the output token length, the average latency increases in both these two scenarios. Moreover, the larger of the output token length, the higher latency increasing ratio is. On the other hand, with the increasing of the output token length, the latency gap between these two scenarios becomes more and more obviously. Especially when the output token length is larger than 1024. However, the latency ratio between these two scenarios remains relatively stable at approximately 1.69.

It is important to note, as indicated in Fig.4 and Fig.5, that even though the performance of distributed LLMs is inferior to that of deploying on a single server, the distributed deployment remains effective when considering that the bandwidth in PCIe is approximately 32 times greater than that available in a distributed environment.

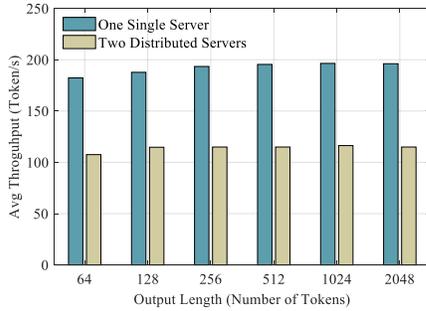

Fig.4. Average generation throughput under different output token length.

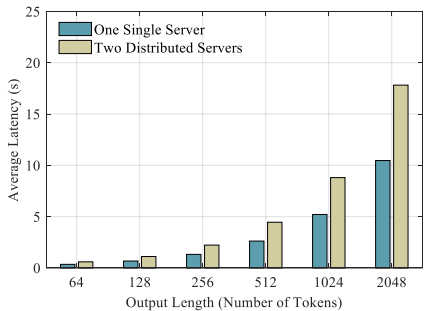

Fig.5. Average latency under different output token length.

In Fig.6, we compare the average latency of distributed deployment under two and three edge servers, respectively. From Fig.6, we can conclude that with the increasing of the input token length, the average latency in both these two scenarios increases. Moreover, the increasing in two distributed servers is more serious than that in three distributed servers. Further, the latency in two distributed servers is higher than that in three distributed servers. Additionally, when the input token length is large, this trend becomes more obviously. For instance, when the input token length is 64, the latency in two-servers scenario is about 1.04 times larger than that in three-servers scenario, this becomes 1.19 times when the input token length is 2048.

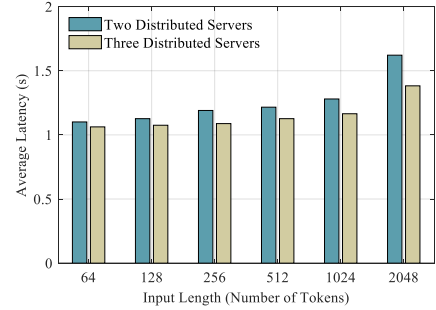

Fig.6. Average latency with different input token length under different number of collaborative servers.

V. CHALLENGES AND POTENTIAL RESEARCH DIRECTIONS

*A. Main Technical Challenges*

**How to Accurately and Timely Discover the Status of edge network?** The status of edge network, including the availability of edge servers and their resources, is crucial for the effective implementation of the proposed CoEL framework, as deployment decisions cannot be made without this information. In practice, edge networks are characterized by high dynamics and heterogeneity due to various factors. For instance, the resources of different edge servers vary and change with network conditions, and user requirements are also diverse and variable. Such information can be captured using existing perception techniques. However, they impose a non-negligible overhead on resource constrained edge networks. Therefore, accurately and timely detecting the status of edge networks with extra controllable overhead is a challenge that needs to be addressed in advance.

**How to precisely predict the distribution of Tokens and the importance/popularity of experts in MoE?** Expert scheduling between different edge servers or between GPU memory and SSD, expert quantification, and token fusion and pruning all rely on this information for adapting to the heterogeneous and dynamic edge resources and user requirements. This information is essential for high-effective LLM deployment at the edge. Consequently, accurately predicting the distribution of tokens and the popularity of experts is vital for achieving high-performance edge LLM deployment.

*B. Potential Research Directions*

Deploying LLMs at the edge is still under developing, there are still numerous issues require further investigation, ranging from architectural considerations to practical applications.

**The multi-dimensions collaborative deployment for**

**LLM**. The CoEL framework currently explores the cooperation between edge servers and the integration of GPU/CPU memory and SSD within the same edge device. In practice, enhancing the performance of edge LLMs can also be achieved through additional collaboration approaches. *Resource collaboration between "cloud-edge-end"*. Based on this collaboration, the efficiency of resource utilization can be significantly enhanced. Consequently, the models, which are larger than current end-LLMs and edge-LLMs, can be deployed on cloud, edge, and end in a distributed and collaborative manner, achieving high effective LLMs. *Model collaboration between small-scale and large-scale LLMs*. In this architecture, the small-scale LLMs are deployed and operated on the resource limited edge or end devices, while the large LLMs is executed on cloud. The knowledge and model parameters are shared between them. Based on this strategy, both the performance of inference accuracy on small-scale LLMs and the capability of generalization on large-scale LLMs can be improved further.

**High-efficient network architecture and technology for distributed and collaborative edge LLMs**. No matter the resource collaboration between edge servers or between cloud, edge, and end, considering the LLM architecture, the distributed and collaborative deployment strategy for edge LLMs involves a significant amount of intermediate data transmission between the distributed devices, such as All-to-All communication. Given that the bandwidth between distributed devices at the edge or end is considerably lower than that of NVlink and InfiniBand, the data transmission delay in edge LLMs is substantially serious compared to cloud data centers. Thus, finding an effective network architecture or technology is crucial for reducing the inference latency of edge LLMs.

**Deployment-aware model training for edge LLMs**. Current LLMs are primarily trained in cloud data centers, which also serve as the environment for model deployment and inference. As a result, the existing model training process does not consider the deployment environment and efficiency. However, for edge LLMs, the heterogeneity and dynamics of devices and resources significantly differ from those in cloud data centers. If the properties of the edge environment can be considered during the model training, such as the load balance between experts, the All-to-All communication between different devices, the token routing strategy, etc., the efficiency of edge LLM deployment could be improved remarkably.

**Online Learning or Incremental Learning for edge LLMs**. As the system operating, model parameters need to be continuously updated to adapt to the new data and environment, ensuring the accuracy and effectiveness of the model. However, updating the entire model at the edge is impractical, as the computing resources and capabilities of edge servers are insufficient for the demands of LLM training. Thus, efficient and lightweight training techniques like incremental learning and online learning may offer potential solutions to this challenge. These methods allow the model to continuously optimize itself by new data and knowledge. This dynamic updating mechanism is crucial for maintaining the relevance and performance of the LLMs in rapidly changing edge environments.

Moreover, considering that deploying LLMs at the edge is still under development, except for the research directions that presented above, there are still many potential interdisciplinary research items are worth to be investigated to improve the performance of edge LLMs further, such as the combination of algorithm and hardware for edge LLMs, the privacy-preserving and security edge LLMs, etc. Additionally, the killer application scenarios of edge LLMs, which is crucial for its wide application, are still being explored currently.

ACKNOWLEDGMENT

This work is supported by fundings from NSFC (no. 62101159), NSF of Shandong (no.ZR2021MF055), and also supported by Hong Kong RGC Areas of Excellence Scheme (AoE/E-601/22-R), Research Impact Fund (No. R5060-19, No. R5034-18), General Research Fund (No. 152203/20E, 152244/21E, 152169/22E, 152228/23E), Key-Area Research and Development Program of Guangdong (No. 2021B0101400003), Shenzhen Science and Technology Innovation Commission (JCYJ20200109142008673).

REFERENCES

[1] J. Achiam, S. Adler, S. Agarwal, L. Ahmad, et al., "GPT-4 Technical Report," arXiv preprint, arXiv: 2303.08774, 2023, pp. 1-100.
[2] H. Touvron, T. Lavril, G. Izacard, X. Martinet, et al., "LLaMA: Open and Efficient Foundation Language Models," arXiv preprint, arXiv: 2302.13971, 2023, pp. 1-27.
[3] A. Q. Jiang, A. Sablayrolles, A. Roux, A. Mensch, et al., "Mixtral of Experts," arXiv preprint, arXiv: 2401.04088, 2024, pp. 1-13.
[4] M. Zhang, J. Cao, X. Shen, Z. Cui, "EdgeShard: Efficient LLM Inference via Collaborative Edge Computing," arXiv preprint, arXiv: 2405.14371, 2024, pp. 1-11.
[5] S. Ye, J. Du, L. Zeng, W. Ou, X. Chu, Y. Lu, X. Chen, "Galaxy: A resource-efficient collaborative edge AI systems for In-situ transformer inference," arXiv Preprint, arXiv: 2405.17245, 2024, pp. 1-10.
[6] J. Xu, Z. Li, W. Chen, Q. Wang, X. Gao, Q. Cai, Z. Ling, "On-device language models: A comprehensive review," arXiv Preprint, arXiv: 2409.00088, 2024 pp. 1-38.
[7] R. Yi, L. Guo, S. Wei, A. Zhou, S. Wang, M. Xu, "Edge-MoE: Fast on-device inference of MoE-based large language models," arXiv preprint, arXiv: 2308.14352, 2023, pp. 1-15.
[8] N. Xue, Y. Sun, Z. Chen, M. Tao, X. Xu, L. Qian, S. Cui, P. Zhang, "WDMoE: Wireless distributed large language models with mixture of experts," arXiv preprint, arXiv: 2405.03131, 2024, pp. 1-6.
[9] S. Minaee, T. Mikolov, N. Nikzad, M. Chenaghlu, R. Socher, X. Amatriain, J. Gao, "Large Language Models: A survey," arXiv preprint, arXiv: 2402, 06196, pp. 1-43.
[10] L. Dai, H. Qi, W. Chen, X. Lu, "High-speed data communication with advanced networks in large language model training," IEEE Micro, vol.44, no.2, 2024, pp. 31-40.
[11] J. Hu, H. Shen, X. Liu, J. Wang, "RDMA transports in datacenter networks: A survey," IEEE Network, vol.38, no.6, 2024, pp: 380-387.
[12] D. Xu, W. Yin, X. Jin, Y. Zhang, S. Wei, M. Xu, X. Liu, "LLMCad: Fast and Scalable on-device Large Language Model Inference," arXiv preprint, arXiv: 2309.04255, 2023, pp. 1-15.
[13] Y. Zheng, Y. Chen, B. Qian, X. Shi, Y. Shu, J. Chen, "A Review on edge large language models: Design, Execution, and Applications," arXiv preprint, arXiv: 2410. 11845, 2024, pp. 1-37.
[14] G. Qu, Q. Chen, W. Wei, Z. Lin, X. Chen, K. Huang, "Mobile Edge Intelligence for Large Language Models: A Contemporary Survey," arXiv preprint, arXiv: 2407.18921, 2024, pp: 1-37.